# Advances in Nanoparticle-Based Targeted Drug Delivery Systems for Colorectal Cancer Therapy: A Review

Mahadi Hasan, Camryn Grace Evett and Jack Burton

Department of Chemistry, University of South Dakota, Vermillion, SD-57069, USA

Corresponding Author: Mahadi Hasan; E-mail: mahadi.hasan01@coyotes.usd.edu

**ABSTRACT:** Colorectal cancer (CRC) continues to be a significant global health burden, prompting the need for more effective and targeted therapeutic strategies. Nanoparticle-based drug delivery systems have emerged as a promising approach to address the limitations of conventional chemotherapy, offering enhanced specificity, reduced systemic toxicity, and improved therapeutic outcomes. This paper provides an in-depth review of the current advancements in the application of nanoparticles as vehicles for targeted drug delivery in CRC therapy. It covers a variety of nanoparticle types, including liposomes, polymeric nanoparticles, dendrimers, and mesoporous silica nanoparticles (MSNs), with a focus on their design, functionalization, and mechanisms of action. This review also examines the challenges associated with the clinical translation of these technologies and explores future directions, emphasizing the potential of nanoparticle-based systems to revolutionize CRC treatment.

---

## 1. Introduction

Colorectal cancer (CRC) is one of the most prevalent malignancies globally, ranking as the third most common cancer and the second leading cause of cancer-related deaths. In 2023 alone, approximately 1.9 million new cases of CRC were diagnosed worldwide, resulting in an estimated 935,000 deaths. The high mortality rate associated with CRC underscores the urgent need for more effective therapeutic strategies, particularly for advanced stages of the disease. Despite significant advancements in surgical techniques and the development of new chemotherapeutic regimens, the prognosis for patients with metastatic CRC remains dismal, with a five-year survival rate of less than 15%.

The challenges associated with conventional chemotherapy in CRC are primarily due to the non-specific nature of these treatments. While chemotherapeutic agents are effective at targeting rapidly dividing cells, their non-selective distribution throughout the body often results in severe side effects, as healthy tissues are also affected. This lack of specificity not only limits the therapeutic efficacy of these drugs but also significantly impacts the patient's quality of life, leading to a dilemma where the treatment itself can become a source of suffering.

To address these challenges, researchers have been focusing on the development of targeted drug delivery systems that aim to deliver therapeutic agents directly to the tumor site while minimizing systemic exposure. Targeted delivery has the potential to enhance drug concentration at the tumor site, reduce off-target effects, and thereby improve the overall efficacy of the treatment. Among the various strategies being explored, nanoparticle-based drug delivery systems have emerged as a particularly promising approach.

Nanoparticles offer several distinct advantages over traditional drug delivery methods. One of the key benefits is the enhanced permeability and retention (EPR) effect, which allows nanoparticles to accumulate preferentially in tumor tissues due to the unique vascular structure of tumors. Additionally, nanoparticles can be engineered to bypass biological barriers, such as the blood-brain barrier, and can be functionalized with specific ligands for active targeting to cancer cells. These characteristics make nanoparticles an attractive vehicle for the delivery of chemotherapeutic agents, particularly in the context of CRC.

This review aims to provide an overview of the latest advancements in nanoparticle-based targeted drug delivery systems for CRC therapy. It will focus on the various types of nanoparticles currently under investigation, their surface modifications to enhance targeting, and the underlying mechanisms by which they achieve tumor-specific delivery. Through this exploration, we hope to shed light on the potential of these systems to revolutionize CRC treatment and improve patient outcomes.

## 2. Nanoparticle Platforms for CRC Therapy

The application of nanoparticles in colorectal cancer (CRC) therapy has gained significant attention due to their potential to improve the specificity and efficacy of drug delivery. Various nanoparticle platforms, including liposomes, polymeric nanoparticles, dendrimers, and mesoporous silica nanoparticles (MSNs), have been extensively studied for their ability to target CRC tumors while minimizing systemic toxicity.

*2.1 Liposomes*

Liposomes, spherical vesicles composed of phospholipid bilayers, are among the most extensively researched nanocarriers for drug delivery, including CRC therapy. Their ability to encapsulate both hydrophilic and hydrophobic drugs, coupled with their biocompatibility, makes them ideal candidates for enhancing drug bioavailability and stability. To prolong circulation time and evade the immune system, liposomes can be surface engineered with polyethylene glycol (PEG) through a process known as PEGylation. This modification has proven effective in increasing the accumulation of liposomes at tumor sites.

Moreover, liposomes can be functionalized by targeting ligands such as antibodies, peptides, or small molecules that specifically bind to overexpressed receptors on CRC cells. For instance, liposomes conjugated with anti-EGFR antibodies have demonstrated preferential accumulation in CRC tumors, leading to enhanced drug delivery and improved therapeutic outcomes. Recent advances have introduced stimuli-responsive liposomes that release their payload in response to specific triggers, such as pH, temperature, or enzymes present in the tumor microenvironment. These "smart" liposomes enable controlled drug release, further reducing off-target effects and enhancing therapeutic efficacy.

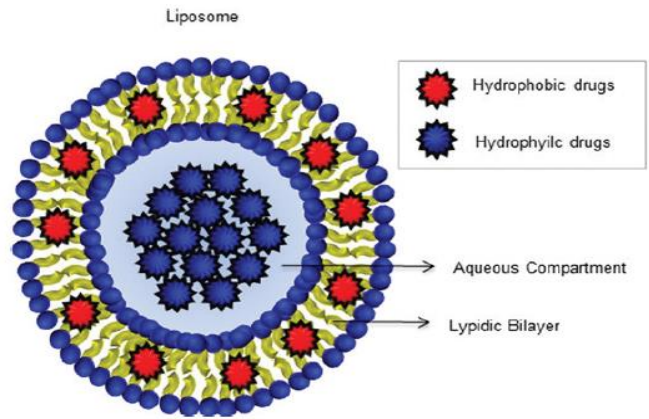

*Figure 1. Liposome drug delivery system. Reproduced with permission from [1]*

*2.2 Polymeric Nanoparticles*

Polymeric nanoparticles, typically composed of biodegradable and biocompatible polymers such as poly(lactic-co-glycolic acid) (PLGA), polylactic acid (PLA), and PEG, represent another promising class of nanocarriers for CRC therapy. These nanoparticles offer controlled drug release through diffusion or polymer matrix degradation. Their versatility in surface modification allows for the attachment of various targeting moieties, facilitating selective delivery to CRC cells.

For example, folate-conjugated polymeric nanoparticles have shown preferential uptake by CRC cells overexpressing the folate receptor, resulting in enhanced drug delivery and reduced systemic toxicity. Additionally, polymeric nanoparticles can be engineered for the co-delivery of multiple therapeutic agents, enabling combination therapy that targets different aspects of tumor growth and survival. The development of stimuli-responsive polymeric nanoparticles, designed to release their cargo in response to the acidic pH or elevated enzyme levels in the tumor microenvironment, has further improved drug accumulation in CRC tumors and demonstrated promising therapeutic outcomes in preclinical models.

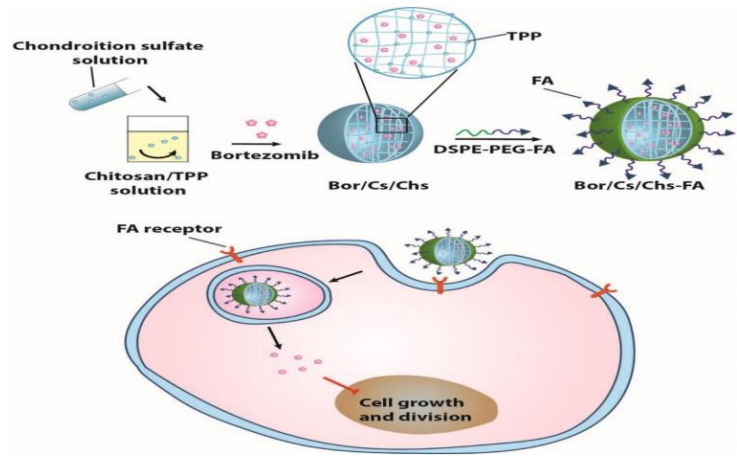

*Figure 2. Schematic illustration of preparing folate targeted polymeric nanoparticles to deliver bortezomib to folate receptor expressing CRCs. Reproduced with permission from [4]*

*2.3 Dendrimers*

Dendrimers are highly branched, monodisperse polymers characterized by a central core, internal branches, and numerous terminal functional groups, which allow for precise control over their size, shape, and surface chemistry. This unique architecture provides a high surface area that can be functionalized with multiple drugs, imaging agents, or targeting ligands, making dendrimers suitable for multifunctional drug delivery systems.

In CRC therapy, dendrimers have been explored for their ability to specifically target cancer cells and deliver chemotherapeutic agents more effectively. For instance, poly(amidoamine) (PAMAM) dendrimers conjugated with folic acid have shown enhanced targeting of CRC cells overexpressing the folate receptor, leading to increased drug delivery and cytotoxicity. Dendrimers can also be engineered to release their payload in response to specific stimuli, such as pH changes or redox conditions, providing a controlled release mechanism that improves the therapeutic index of the delivered drug.

Beyond drug delivery, dendrimers hold potential as imaging agents for tumor diagnosis and monitoring treatment response. For example, gadolinium-labeled dendrimers have been developed for magnetic resonance imaging (MRI) of CRC tumors, providing high-resolution images that aid in early detection and characterization of the disease.

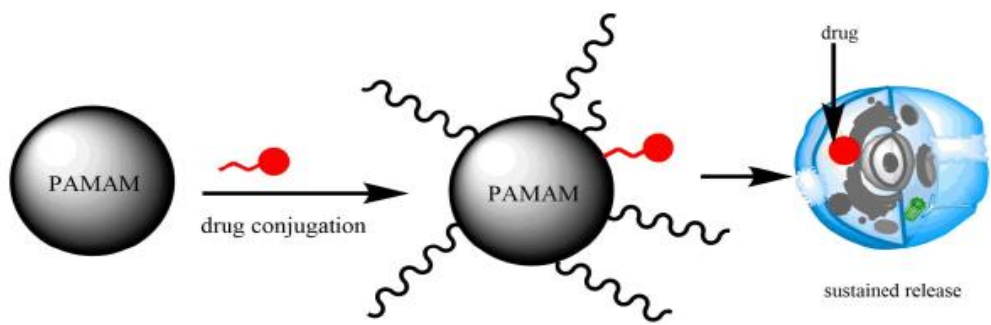

*Figure 3. Schematic representation of PAMAM dendrimer drug conjugate. Reproduced with permission from [3]*

2.4 Mesoporous Silica Nanoparticles

Mesoporous silica nanoparticles (MSNs) have emerged as a highly promising class of nanocarriers for CRC therapy due to their unique structural features, including a high surface area, large pore volume, and tunable pore size. These characteristics allow for the efficient loading of a wide range of therapeutic agents, which can be encapsulated within the mesopores to protect them from degradation and provide a sustained release profile.

A significant advantage of MSNs is their ease of surface functionalization, which can be utilized to attach targeting ligands, imaging agents, or stimuli-responsive gatekeepers. For example, MSNs functionalized with hyaluronic acid (HA) have shown preferential binding to CRC cells overexpressing the CD44 receptor, resulting in enhanced cellular uptake and improved therapeutic efficacy. Moreover, MSNs can be designed to release their cargo in response to specific triggers in the tumor microenvironment, such as acidic pH, reducing conditions, or enzymatic activity.

Recent studies have explored the potential of MSNs for the co-delivery of chemotherapeutic agents and gene therapies in CRC. For instance, MSNs loaded with doxorubicin and a siRNA targeting the anti-apoptotic protein Bcl-2 have shown synergistic effects in inducing apoptosis in CRC cells. This multifunctional approach not only enhances therapeutic efficacy but also addresses drug resistance, a significant challenge in CRC treatment.

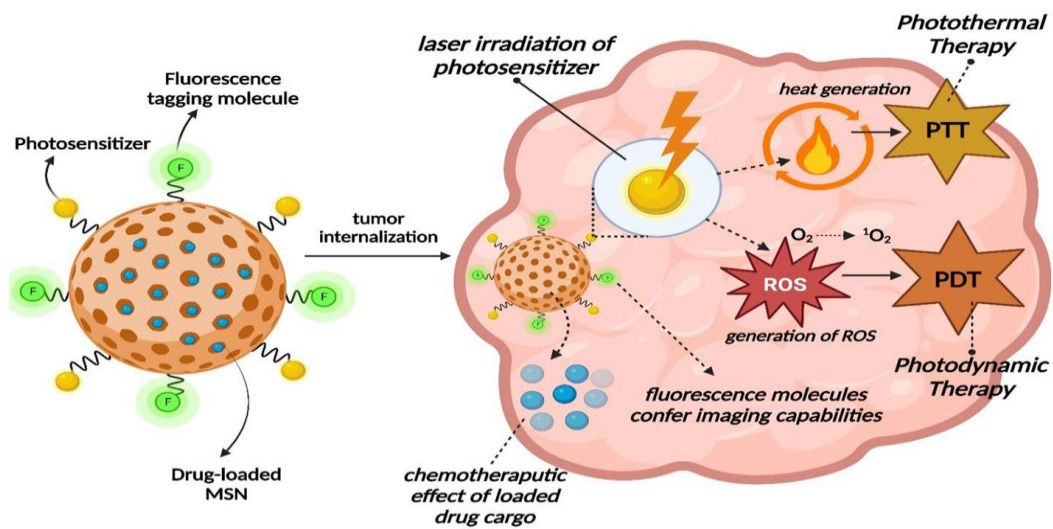

Figure 4. Graphical representation of a drug-loaded theranostic MSN platform designed to have light -based triggers for therapeutic responses. Reproduced with permission from [2]

### 3. Functionalization Strategies for Targeted Drug Delivery

The effectiveness of nanoparticle-based drug delivery systems in colorectal cancer (CRC) therapy is significantly influenced by their ability to selectively target tumor cells while minimizing impact on healthy tissues. The functionalization of nanoparticles with targeting ligands is a critical approach to achieve this selectivity. By attaching various ligands, including antibodies, peptides, small molecules, and aptamers, to the nanoparticle surface, researchers can enhance the recognition and binding of these nanoparticles to specific receptors that are overexpressed on CRC cells, thereby improving therapeutic efficacy and reducing systemic toxicity.

*3.1 Active Targeting*

Active targeting is a sophisticated strategy that involves modifying nanoparticles with ligands that specifically bind to receptors on the surface of CRC cells. This targeting approach not only increases the accumulation of nanoparticles at the tumor site but also facilitates receptor-mediated endocytosis, thus enhancing the intracellular delivery of the therapeutic agent.

One of the most thoroughly studied receptors in CRC is the epidermal growth factor receptor (EGFR), which is overexpressed in a significant proportion of CRC cases. Nanoparticles functionalized with anti-EGFR antibodies have shown enhanced binding and internalization within CRC cells, resulting in increased drug delivery to the tumor and improved therapeutic outcomes. For instance, in preclinical models, EGFR-targeted nanoparticles demonstrated a significant increase in drug accumulation in CRC tumors compared to non-targeted nanoparticles, leading to more effective tumor regression.

Similarly, folate receptors, which are overexpressed in many CRC cases, have been exploited as a target for drug delivery. Nanoparticles conjugated with folic acid have been developed to exploit this overexpression, leading to preferential uptake by CRC cells and reducing off-target effects. Studies have shown that folate-functionalized nanoparticles exhibit significantly higher uptake in CRC cells compared to non-functionalized nanoparticles, which translates to better therapeutic outcomes in vivo.

Beyond EGFR and folate receptors, other promising targets for active targeting in CRC include the transferrin receptor, integrins, and the vascular endothelial growth factor (VEGF) receptor [43]. For example, nanoparticles functionalized with ligands targeting the transferrin receptor have demonstrated enhanced specificity for CRC cells, which express higher levels of this receptor compared to normal cells. Similarly, integrin-targeted nanoparticles have shown improved binding and internalization in CRC cells, leading to enhanced drug delivery and reduced systemic toxicity. VEGF receptor-targeted nanoparticles have also been explored, given the role of VEGF in tumor angiogenesis, with studies showing that such targeted nanoparticles can inhibit tumor growth by blocking angiogenesis and directly delivering cytotoxic agents to the tumor.

The versatility of active targeting allows for the customization of nanoparticles to target multiple receptors simultaneously, which can enhance specificity and reduce the likelihood of drug resistance. By combining ligands for different receptors, researchers can create multifunctional nanoparticles that offer superior therapeutic outcomes in CRC therapy.

*3.2 Passive Targeting*

Passive targeting is another important strategy that exploits the enhanced permeability and retention (EPR) effect, a phenomenon whereby nanoparticles preferentially accumulate in tumor tissues due to the leaky vasculature and poor lymphatic drainage that are characteristic of solid tumors. Although passive targeting does not involve specific ligand-receptor interactions, it allows for the preferential concentration of nanoparticles in the tumor microenvironment, thereby increasing the local concentration of the drug and enhancing its efficacy.

While passive targeting is inherently less selective than active targeting, it plays a crucial role in delivering nanoparticles to solid tumors, including CRC. The EPR effect is particularly effective for larger nanoparticles, such as liposomes and polymeric nanoparticles, which can accumulate in tumor tissues more readily than smaller particles. However, the effectiveness of the EPR effect can vary significantly between patients and tumor types, leading to variability in therapeutic outcomes. For example, tumors with dense stromal tissue or poor vasculature may exhibit a reduced EPR effect, limiting the accumulation of nanoparticles.

To address these limitations, combining passive and active targeting strategies has been proposed as a synergistic approach that can enhance the specificity and efficacy of nanoparticle-based drug delivery systems. For instance, nanoparticles that initially accumulate in the tumor via the EPR effect can be further targeted to specific cancer cells using surface ligands, thus providing a dual-targeting approach that maximizes drug delivery and minimizes off-target effects. This dual-targeting strategy has been shown to improve therapeutic outcomes in preclinical models of CRC by ensuring that the drug is delivered directly to the tumor cells, even in heterogeneous tumor environments.

Overall, the integration of active and passive targeting strategies represents a promising approach for improving the precision and effectiveness of nanoparticle-based therapies for CRC. As research continues to advance, the development of more sophisticated targeting mechanisms will likely lead to even greater improvements in the treatment of this challenging disease.

## 4. Challenges in Nanoparticle-Based CRC Therapy

Nanoparticle-based drug delivery systems offer significant promise for improving colorectal cancer (CRC) therapy by enhancing the precision of drug delivery, reducing systemic toxicity, and overcoming limitations of conventional chemotherapy. However, several critical challenges must be addressed to fully realize the potential of these systems and facilitate their successful clinical translation.

*4.1 Scalability and Manufacturing*

One of the foremost challenges in developing nanoparticle-based therapies is the scalability of production. Manufacturing nanoparticles with consistent quality, size, and surface properties on a large scale is essential for clinical application. Variability in nanoparticle synthesis can significantly impact drug loading, release profiles, and targeting efficacy, ultimately affecting therapeutic outcomes. For instance, microfluidic technologies have been explored to address scalability issues by allowing precise control over nanoparticle size and uniformity, which is crucial for consistent drug delivery performance.

Achieving uniformity in nanoparticle production requires the development of robust manufacturing processes. Quality control measures during the manufacturing process are essential to ensure batch-to-batch consistency. Overcoming these scalability challenges is critical for the widespread adoption of nanoparticle-based therapies in clinical settings.

*4.2 Stability and Shelf-Life*

The stability of nanoparticles during storage, transportation, and administration is another major challenge that must be addressed to ensure their efficacy. Nanoparticles must maintain their physicochemical properties, such as size, surface charge, and drug encapsulation, throughout their shelf-life. Factors such as aggregation, drug leakage, and degradation of the nanoparticle matrix can compromise the performance of the drug delivery system.

For example, nanoparticle aggregation can lead to increased particle size, which may alter biodistribution and reduce tumor accumulation. Drug leakage during storage can reduce the effective dose delivered to the tumor, diminishing therapeutic efficacy. To address these challenges, researchers are developing more stable nanoparticle formulations, including the use of stabilizing agents such as PEG and cryoprotectants that prevent aggregation and degradation.

*4.3 Immunogenicity and Toxicity*

The immunogenicity and potential toxicity of nanoparticles are significant concerns that must be carefully evaluated before clinical use. The immune system can recognize and clear nanoparticles from circulation, reducing their effectiveness and potentially causing adverse reactions. The rapid clearance of nanoparticles by the mononuclear phagocyte system (MPS) can limit their circulation time and prevent them from reaching the tumor site in sufficient concentrations.

To mitigate these issues, several strategies have been developed to reduce the immunogenicity and toxicity of nanoparticles. PEGylation, the attachment of polyethylene glycol (PEG) chains to the nanoparticle surface, is a widely used approach to evade immune recognition and extend circulation time. Additionally, biodegradable polymers such as poly(lactic-co-glycolic acid) (PLGA) are used to ensure that nanoparticles break down into non-toxic byproducts after delivering their therapeutic payload.

*4.4 Tumor Heterogeneity and Microenvironment*

The heterogeneity of tumors and the dynamic nature of the tumor microenvironment pose significant challenges for targeted drug delivery. CRC tumors are characterized by considerable heterogeneity in receptor expression, extracellular matrix composition, and vascularization, which can affect nanoparticle distribution and efficacy.

For instance, differences in receptor expression levels between tumor cells can lead to unequal uptake of targeted nanoparticles, reducing overall therapeutic efficacy. The dense extracellular matrix in some tumors can act as a physical barrier, preventing nanoparticles from penetrating deep into the tumor tissue. Additionally, variations in tumor vascularization can result in uneven nanoparticle distribution, with some areas receiving higher drug concentrations than others.

To address these challenges, researchers are developing personalized and adaptive nanoparticle-based therapies that account for the unique characteristics of each patient's tumor. Techniques such as imaging-guided drug delivery, which uses real-time imaging to monitor nanoparticle distribution within the tumor, are being explored to optimize dosing and improve treatment outcomes. Additionally, multifunctional nanoparticles that target multiple pathways or receptors simultaneously are being developed to overcome the limitations posed by tumor heterogeneity.

## 5. Future Directions

The future of nanoparticle-based drug delivery in CRC therapy lies in the development of more sophisticated and personalized delivery systems. Several promising research directions include:

*5.1 Multifunctional Nanoparticles*

The development of multifunctional nanoparticles that combine drug delivery, imaging, and diagnostic capabilities offers a promising approach to improving CRC therapy. These theranostic nanoparticles can provide real-time monitoring of drug distribution, release, and therapeutic response, enabling personalized treatment strategies.

*5.2 Nanoparticles for Combination Therapy*

Nanoparticles designed for the co-delivery of multiple therapeutic agents, such as chemotherapeutics, gene therapies, and immunotherapies, hold significant potential for overcoming drug resistance and improving therapeutic outcomes. Combination therapy approaches can target different pathways involved in tumor growth and survival, providing a more comprehensive treatment strategy.

*5.3 Patient-Specific Nanoparticles*

The use of patient-specific tumor models and advanced imaging techniques can aid in the design of personalized nanoparticle-based therapies. By tailoring the nanoparticle properties, drug loading, and targeting ligands to the specific characteristics of the patient's tumor, it is possible to maximize therapeutic efficacy and minimize adverse effects.

## 6. Conclusion

Nanoparticles represent a promising platform for the targeted delivery of chemotherapeutic agents in colorectal cancer therapy. The ability to functionalize nanoparticles with targeting ligands, along with the development of stimuli-responsive systems, has the potential to significantly improve the specificity and efficacy of CRC treatment. However, several challenges must be addressed to translate these technologies from the laboratory to the clinic. Ongoing research in the field is expected to overcome these challenges, paving the way for the development of next-generation nanomedicines that can revolutionize CRC therapy.